\documentclass[fleqn,usenatbib]{mnras}
\usepackage{newtxtext,newtxmath}
\usepackage[T1]{fontenc}
\usepackage{ae,aecompl}

\usepackage{color}\usepackage{graphicx}
\usepackage{amsmath}	
\usepackage{amssymb}	
\usepackage[figuresleft]{rotating}
\newcommand{\HI}{H{\,\small I}}

\newcommand{\ltsima} {$\; \buildrel < \over \sim \;$}
\newcommand{\gtsima} {$\; \buildrel > \over \sim \;$}
\newcommand{\lta} {\lower.5ex\hbox{\ltsima}}
\newcommand{\kmsMp}{km s$^{-1}$\,Mpc$^{-1}$}
\newcommand{\gta} {\lower.5ex\hbox{\gtsima}}

\newcommand{\kms}{km\,s$^{-1}$}

\newcommand{\cip}{[C\,{\sc i}]~$^{3}$P$_{1}$-$^{3}$P$_{0}$}
\newcommand{\ci}{[C\,{\sc i}]}

\title[Carbon recycling in the cold CGM of the Spiderweb]{Giant galaxy growing from recycled gas: ALMA maps the circumgalactic molecular medium of the Spiderweb in [C\,{\Large I}]}
\author[B.\,H.\,C. Emonts et al.]{B.\,H.\,C. Emonts$^{1}$, M.\,D. Lehnert$^{2}$, H. Dannerbauer$^{3,4}$, C.\,De Breuck$^{5}$, 
\newauthor M. Villar-Mart\'{i}n$^{6}$, G.\,K. Miley$^{7}$, J.\,R. Allison$^{8,9}$, B. Gullberg$^{10}$,  
\newauthor N.\,A. Hatch$^{11}$, P. Guillard$^{2}$, M.\,Y. Mao$^{12}$, R.\,P. Norris$^{13,14}$\\
$^{1}$National Radio Astronomy Observatory, 520 Edgemont Road, Charlottesville, VA 22903\\
$^{2}$Sorbonne Universit\'{e}, CNRS, UMR 7095, Institut d'Astrophysique de Paris, 98bis bvd Arago, 75014, Paris, France\\
$^{3}$Instituto de Astrof\'{i}sica de Canarias, E-38205 La Laguna, Tenerife, Spain\\
$^{4}$Universidad de La Laguna, Dpto. Astrof\'{i}sica, E-38206 La Laguna, Tenerife, Spain\\
$^{5}$European Southern Observatory, Karl Schwarzschild Strasse 2, 85748 Garching, Germany\\
$^{6}$Centro de Astrobiolog\'{i}a (INTA-CSIC), Ctra de Torrej\'{o}n a Ajalvir, km 4, 28850 Torrej\'{o}n de Ardoz, Madrid, Spain\\
$^{7}$Leiden Observatory, University of Leiden, P.O. Box 9513, 2300 RA Leiden, Netherlands\\
$^{8}$Sydney Institute for Astronomy, School of Physics A28, The University of Sydney, NSW 2006, Australia\\
$^{9}$ARC Centre of Excellence for All-sky Astrophysics in 3 Dimensions (ASTRO 3D)\\
$^{10}$Centre for Extragalactic Astronomy, Department of Physics, Durham University, Durham DH1 3LE, UK\\
$^{11}$School of Physics and Astronomy, University of Nottingham, University Park, Nottingham NG7 2RD, UK\\
$^{12}$Jodrell Bank Observatory, University of Manchester, Macclesfield, Cheshire SK11 9DL, UK\\
$^{13}$CSIRO Astronomy and Space Science, Australia Telescope National Facility, PO Box 76, Epping NSW, 1710, Australia\\
$^{14}$Western Sydney University, Penrith South, NSW 1797, Australia}
\begin{document}

\date{}

\maketitle

\begin{abstract}
The circumgalactic medium (CGM) of the massive Spiderweb Galaxy, a conglomerate of merging proto-cluster galaxies at z=2.2, forms an enriched interface where feedback and recycling act on accreted gas.
This is shown by observations
of \ci, CO(1-0) and CO(4-3) performed with the Atacama Large Millimeter Array (ALMA) and Australia Telescope Compact Array (ATCA). \ci\ and CO(4-3) are detected across $\sim$50 kpc, following the distribution of previously detected low-surface-brightness CO(1-0) across the CGM. This confirms our previous results on the presence of a cold molecular halo. The central radio galaxy MRC\,1138-262 shows a very high global $L^{\prime}${\scriptsize CO(4-3)}/$L^{\prime}${\scriptsize CO(1-0)} $\sim$ 1, suggesting that mechanisms other than FUV-heating by star formation prevail at the heart of the Spiderweb Galaxy. Contrary, the CGM has $L^{\prime}${\scriptsize CO(4-3)}/$L^{\prime}${\scriptsize CO(1-0)} and $L^{\prime}${\scriptsize [CI]}/$L^{\prime}${\scriptsize CO(1-0)} similar to the ISM of five galaxies in the wider proto-cluster, and its carbon abundance, $X${\scriptsize [CI]}/$X${\scriptsize H$_2$}, resembles that of the Milky Way and starforming galaxies. The molecular CGM is thus metal-rich and not diffuse, confirming a link between the cold gas and \textit{in-situ} star formation. Thus, the Spiderweb Galaxy grows not directly through accretion of gas from the cosmic web, but from recycled gas in the GCM. 
\end{abstract}


\section{Introduction}
\label{sec:intro}

Most of the baryons in the Universe lie outside galaxies. We can study baryonic halos around galaxies through absorption lines towards distant quasars,~or cooling-radiation emitted in Ly$\alpha$. Absorption-line studies detect $\sim$100\,kpc halos of warm, T$\sim$10$^{4}$\,K, enrich gas around high-$z$ galaxies and quasars \citep[e.g.,][]{pro14,nee17}. 

We recently discovered that the coldest gas phase can also exist in such environments, by revealing a molecular gas reservoir across the halo of the massive forming Spiderweb Galaxy at z\,=\,2.2 \citep[][hereafter EM16]{emo16}. The Spiderweb Galaxy is a conglomerate of starforming galaxies that surround the radio galaxy MRC\,1138-262, and that are embedded in a giant Ly$\alpha$ halo \citep{pen97,mil06}. We refer to the entire 200\,kpc region of the Ly$\alpha$ halo as the ``Spiderweb Galaxy'', because it will likely evolve into a single dominant cluster galaxy \citep{hat09}. The Spiderweb Galaxy is part of a larger proto-cluster \citep[][]{kur04,kod07,dan14}.

The halo gas spans a wide range of temperatures and densities (T\,$\sim$\,100-10$^{7}$\,K,\,n\,$\sim$\,10$^{-3}$-10$^{4}$\,cm$^{-3}$; \citealt{car02}, EM16). Across the inner $\sim$70 kpc, we detected $\sim$10$^{11}$ M$_{\odot}$ of molecular gas via CO(1-0) (EM16). The location and velocity of the CO, as well as its large angular scale (EM16, their Fig. S1), imply that the bulk of the molecular gas is found in the gaseous medium that lies between the brightest galaxies in the halo. We refer to this gaseous medium as the circumgalactic medium (CGM). There is also diffuse blue light across the halo, indicating that \textit{in-situ} star formation occurs within the CGM \citep{hat08}. Since the surface densities of the molecular gas and the rate of star formation fall along the Schmidt-Kennicutt relation, the CO(1-0) results provided the first direct link between star formation and cold molecular gas in the CGM of a forming massive galaxy at high-$z$ (EM16). Extended CO is also found in the CGM of a massive galaxy at $z$\,=\,3.47 \citep{gin17}.

Here we present observations sensitive to low-surface-brightness extended emission of atomic carbon, \cip\ (hereafter \ci) in the CGM of the Spiderweb Galaxy. We supplement these with observations of CO(1-0) and CO(4-3) to study the chemical composition and excitation conditions of the gas. \ci\ and CO(1-0) are fully concomitant in molecular clouds across the Milky Way \citep{ojh01,ike02}. They have a similar critical density, with the \ci\ $J$=1 level well populated down to T$_{\rm k}$\,$\sim$\,15\,K \citep{pap04}. A large positive K-correction means that, at comparable resolution, \ci\ is much brighter than CO(1-0). This becomes progressively more advantageous towards higher redshifts, as the instrumental T$_{\rm sys}$ at the corresponding frequencies become more comparable \citep{pap04,tom14}. Furthermore, a high cosmic ray flux from star formation or radio jets may reduce the CO abundance in favor of \ci\ \citep{bis15,bis17}.

We assume H$_{0} = 71$\,\kmsMp, $\Omega_\textrm{M} = 0.27$ and $\Omega_{\Lambda} = 0.73$, i.e., 8.4 kpc/$^{\prime\prime}$ and $D_{\rm L} = 17309$ Mpc at z\,=\,2.2 (EM16).

\section{Observations}
\label{sec:observations}

We observed the Spiderweb for 1.8~hrs on-source during ALMA cycle-3 on 16 Jan 2016 in its most compact 12m configuration (C36-1; baselines 15$-$161m). We placed two adjacent spectral windows of 1.875 GHz on \cip\ at $\nu_{\rm obs}$\,$\sim$\,155.7 GHz ($\nu_{\rm rest}$\,=\,492.16 GHz) and another two on CO(4-3) at $\nu_{\rm obs}$\,$\sim$\,145.8 GHz ($\nu_{\rm rest}$\,=\,461.04 GHz). The ALMA data were reduced in CASA (Common Astronomy Software Applications; \citealt{mcm07}). We binned the data to 30 \kms\ channels and cleaned signal $\geq$1 mJy\,bm$^{-1}$. We then Hanning smoothed the data to a resolution of 60 \kms. The resulting noise is 0.085 mJy\,beam$^{-1}$ channel$^{-1}$. We imaged our field using natural weighting out to $\sim$33$^{\prime\prime}$, where the primary beam response drops to $\sim$10$\%$ sensitivity. The synthesized beam is 2.3$^{\prime\prime}$\,$\times$1.5$^{\prime\prime}$ with PA=73.4$^{\circ}$. 

Using the ATCA, we made a 2-pointing mosaic with an on-source time of $\sim$90\,hrs per pointing at $\nu_{\rm obs}$\,$\sim$\,36.5 GHz to observe CO(1-0). The first pointing was centred on the Spiderweb Galaxy (EM16). The second one was centred $\sim$23$^{\prime\prime}$ to the west and observed in Jan 2016 in 750C configuration for 42\,hrs and in April 2016 in H214 configuration for 40\,hrs on-source. The observing strategy and data reduction in MIRIAD followed EM16. Because the Spiderweb Galaxy was located near the edge of the primary beam in the second pointing, its strong continuum caused beam-smearing errors that could not be completely eliminated, even with model-based continuum subtraction in the ($u$,$v$)-domain \citep{all12}. This prevented us from improving the image of the faint CO(1-0) in the halo compared with EM16. We imaged both pointings separately using natural weighting, and combined them using the task LINMOS. We binned the channels to 34 \kms\ and applied a Hanning smooth, resulting in a resolution of 68 \kms. The noise in the center of the mosaic is 0.073 mJy\,bm$^{-1}$, with a beam of 4.7$^{\prime\prime}$\,$\times$\,4.1$^{\prime\prime}$ (PA\,36.3$^{\circ}$). Velocities are in the optical frame relative to $z$=2.1612.

\vspace{-1mm}
\section{Results}
\label{sec:results}

\begin{figure*}
\centering
\includegraphics[width=0.82\textwidth]{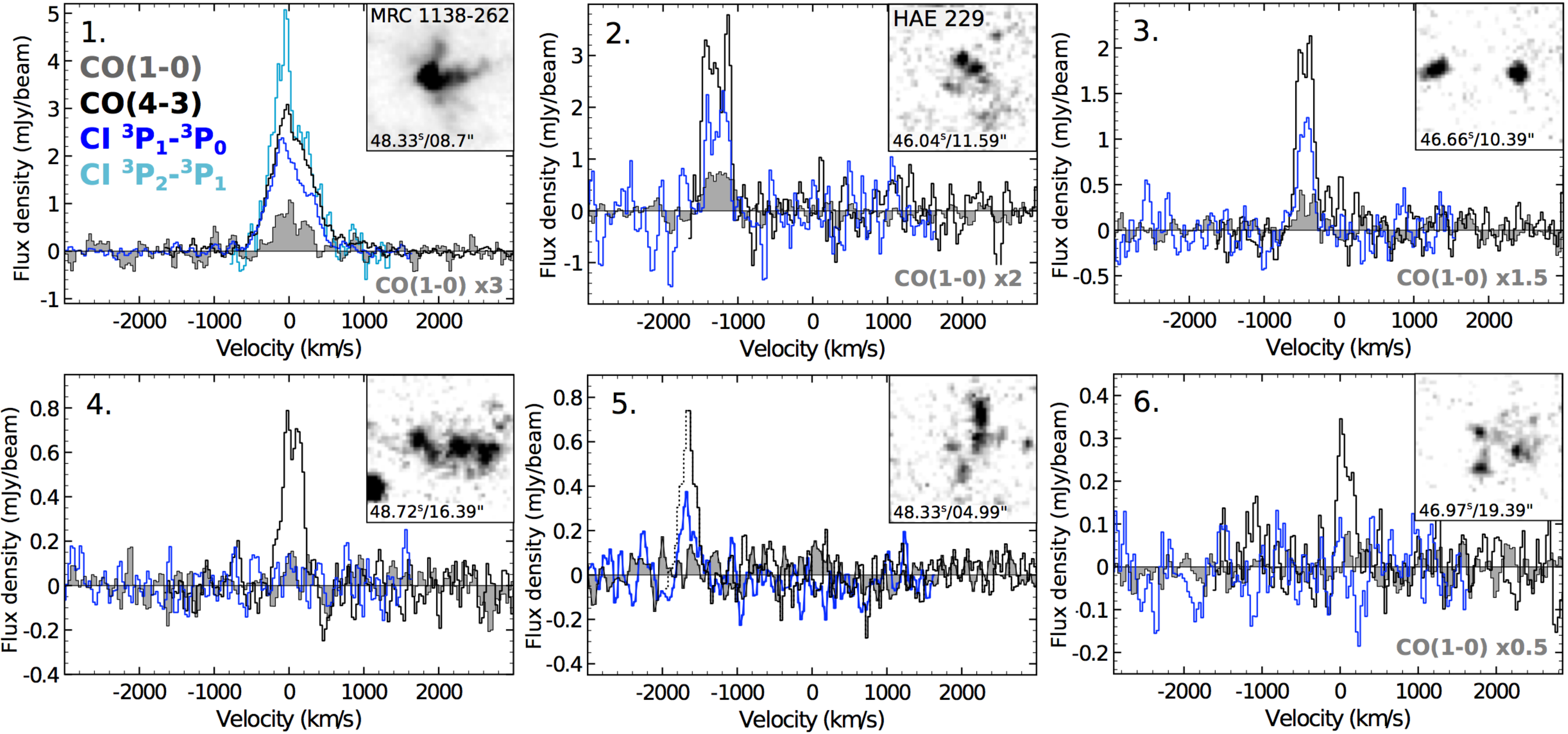}
\caption{Spectra of \cip (blue), CO(4-3) (black), and CO(1-0) (grey) in the six proto-cluster galaxies. Except for radio galaxy MRC\,1138-262, all are located far outside the Ly$\alpha$ halo of the Spiderweb Galaxy. Some of the CO(1-0) spectra are scaled up by a factor indicated at the bottom-right, to better visualize them. For MRC\,1138-262 we also show the [C\,{\sc i}]~$^{3}$P$_{2}$-$^{3}$P$_{1}$ line (light blue), derived by tapering and smoothing the ALMA data from \citet{gul16} to the resolution of our \cip\ data. Galaxy $\#$2 is H$\alpha$ emitter HAE\,229, for which \citet{dan17} detected CO(1-0) across a large disk. The CO(4-3) line of galaxy $\#$5 fell at the edge of the band, and the dotted line estimates the profile if it is symmetric. The top-right inset in each panel shows a 2$^{\prime\prime}$\,$\times$\,2$^{\prime\prime}$ region of the galaxy in {\it HST}/ACS F475W+F814W imaging \citep{mil06}. Coordinates in seconds and arcsec are relative to RA=11h40m and $\delta$=$-$26$^{\circ}$29$^{\prime}$.}
\label{fig:galaxies}
\end{figure*}

Fig.~\ref{fig:galaxies} shows \ci, CO(4-3) and CO(1-0) from proto-cluster galaxies within 250\,kpc radius around the Spiderweb Galaxy. Six proto-cluster galaxies show line emission (Table\,\ref{tab:results}).

The central radio galaxy MRC\,1138-262 is covered fully by the ALMA beam and shows an extraordinary high global $L^{\prime}_{\rm CO(4-3)}$/$L^{\prime}_{\rm CO(1-0)}$\,$\sim$\,1 (Table\,\ref{tab:results}). In metal-rich environments, such high global gas excitation states are hard to achieve with far-UV photons from star formation, and cloud-heating mechanisms due to cosmic rays, jet-induced shocks, or gas turbulence must be prevailing \citep{pap08,pap12,ivi12}. MRC\,1138-262 also has a high $L^{\prime}_\textrm{[CI]}$/$L^{\prime}_\textrm{CO}$$\sim$0.67, exceeding that of most submm galaxies (SMGs), quasi-stellar objects (QSOs), and lensed galaxies \citep{wal11, ala13, bot17}. We compare our \cip\ detection with [C\,{\sc i}]~$^{3}$P$_{2}$-$^{3}$P$_{1}$ data from \citet{gul16}, which we tapered and smoothed to the same spatial resolution (Fig.~\ref{fig:galaxies}). We derive a \ci\ fine-structure ratio of $L^{\prime}_\mathrm{[CI]2 \rightarrow 1}$/$L^{\prime}_\mathrm{[CI]1 \rightarrow 0}$\,$\sim$\,0.62, which implies an excitation temperature  T$_\mathrm{ex}$$\sim$32\,K for optically thin gas \citep{stu97}.

\begin{table*}
\caption{Emission-line properties. Velocity $v$ is relative to $z$=2.1612, while $v$ and FWHM are derived by fitting a Gaussian function to the CO(4-3) line (\ci\ for galaxy $\#$5). The ratios of the brightness luminosity ($L'$) are r$_\mathrm{CI/1-0}$ = $L^{\prime}_\mathrm{[CI]1 \rightarrow 0}$/$L^{\prime}_\mathrm{CO(1-0)}$, r$_\mathrm{CI/4-3}$ = $L^{\prime}_\mathrm{[CI]1 \rightarrow 0}$/$L^{\prime}_{\rm CO(4-3)}$, and r$_{\rm 4-3/1-0}$ = $L^{\prime}_{\rm CO(4-3)}$/$L^{\prime}_{\rm CO(1-0)}$. The molecular gas mass M$_\mathrm{H_2}$ is derived from $L'_{\rm CO(1-0)}$ \citep{sol05}, assuming $\alpha_{\rm CO}$\,=\,M$_{\rm H_2}$/$L'_{\rm CO(1-0)}$\,=\,0.8 M$_{\odot}$ (K \kms\ pc$^{2}$)$^{-1}$ for galaxies $\#$1$-$6, and $\alpha_{\rm CO}$\,=\,4 M$_{\odot}$ (K \kms\ pc$^{2}$)$^{-1}$ for the CGM (see EM16). The \ci\ mass (M$_{\rm [CI]}$) is estimated following \citet{wei05}, assuming $T_{\rm ex}$\,=\,30\,K. Errors (in brackets) include uncertainties in $I$ from both the noise (see Eqn. 2 of \citealt{emo14}; also \citealt{sag90}) and the absolute flux calibration (5$\%$ for ALMA; 20$\%$ for ATCA).}
\label{tab:results}
\begin{tabular}{lcccccccccc}
$\#$ & $v$ & FWHM & $I_{\rm [CI]1 \rightarrow 0}$ & $I_{\rm CO(1-0)}$ & $I_{\rm CO(4-3)}$ & r$_{\rm CI/1-0}$ & r$_{\rm CI/4-3}$ & r$_{\rm 4-3/1-0}$ & M$_{\rm H_2}$ & M$_{\rm [CI]}$ \\
 & km/s & km/s & \multicolumn{3}{c}{Jy/bm$\times$km/s} & & & & 10$^{10}$\,M$_{\odot}$ & 10$^{6}$\,M$_{\odot}$  \\
\hline
1 & $-$ & 610\,(10) &1.32\,(0.07)& 0.11\,(0.03)$^{\dagger}$ & 1.77\,(0.09) & 0.66\,(0.18) & 0.67\,(0.04) & 1.00\,(0.28) & 2.0\,(0.6) & 21\,(1.0) \\
2 & -1290\,(10) & 375\,(20) & 0.60\,(0.08) & 0.15\,(0.04)$^{\ddagger}$ & 1.19\,(0.08) & 0.22\,(0.05) & 0.44\,(0.07) & 0.49\,(0.17) &  2.8\,(0.6) & 9.5\,(1.2) \\
3 & -450\,(10)  & 255\,(15) & 0.29\,(0.03) & 0.06\,(0.02) & 0.54\,(0.03) & 0.27\,(0.09) & 0.47\,(0.05) & 0.56\,(0.19) & 1.1\,(0.4) & 4.6\,(0.5) \\
4 & 45\,(10) & 310\,(20) & $<$0.03 &  $<$0.03 & 0.25\,(0.02) & $-$ & $<$0.11 & $>$0.52 & $<$0.6 & $<$0.5 \\
5 & -1655\,(10) & 220\,(15) & 0.08\,(0.01) & $<$0.03 & 0.16\,(0.01)$^{*}$ & $>$0.15 & 0.44\,(0.06) & $>$0.33 & $<$0.6 & 1.2\,(0.1)  \\
6 & 60\,(15) & 200\,(25) & $<$0.04 & $<$0.03 & 0.13\,(0.02) & $-$ & $<$0.27  & $>$0.27 & $<$0.6 & $<$0.6 \\
\multicolumn{2}{l}{CGM \hspace{3mm} $-$} & $-$ & 0.56\,(0.09) & 0.11\,(0.04) & 0.79\,(0.07) & 0.28\,(0.11) & 0.62\,(0.11) & 0.45\,(0.17) & 10\,(4) & 8.9\,(1.4)\\
\hline
\end{tabular} 
\flushleft 
$^{\dagger}$ The ATCA profile in Fig.\,\ref{fig:galaxies} provides an upper limit to the CO(1-0) content of MRC\,1138-262, because the CO(1-0) data have a larger beam than the \ci\ and CO(4-3) data, and therefore include more extended CO(1-0). The corresponding $I_{\rm CO(1-0)}$\,$\la$\,0.126 Jy\,bm$^{-1}$ $\times$ \kms. To estimate lower limit values, we tapered existing high-resolution VLA data (EM16) to the spatial resolution of our ALMA data. This gives $I_{\rm CO(1-0)}$\,$\ga$\,0.101 Jy\,bm$^{-1}$\,$\times$\,\kms. However, these VLA data have lower sensitivity, hence likely underestimate the full width of the profile. $I_{\rm CO(1-0)}$ in the Table is a weighted average of the two values, although both values are within the uncertainties.\\
$^{\ddagger}$ \citet{dan17} previously reported a somewhat higher $I_{\rm CO(1-0)}$, although our estimate agrees to within the uncertainties.\\
$^{*}$ CO(4-3) falls at the edge of the band and half the profile is missing. We derive values assuming that the line profile is symmetric.
\end{table*} 

\begin{figure*}
\centering
\includegraphics[width=0.78\textwidth]{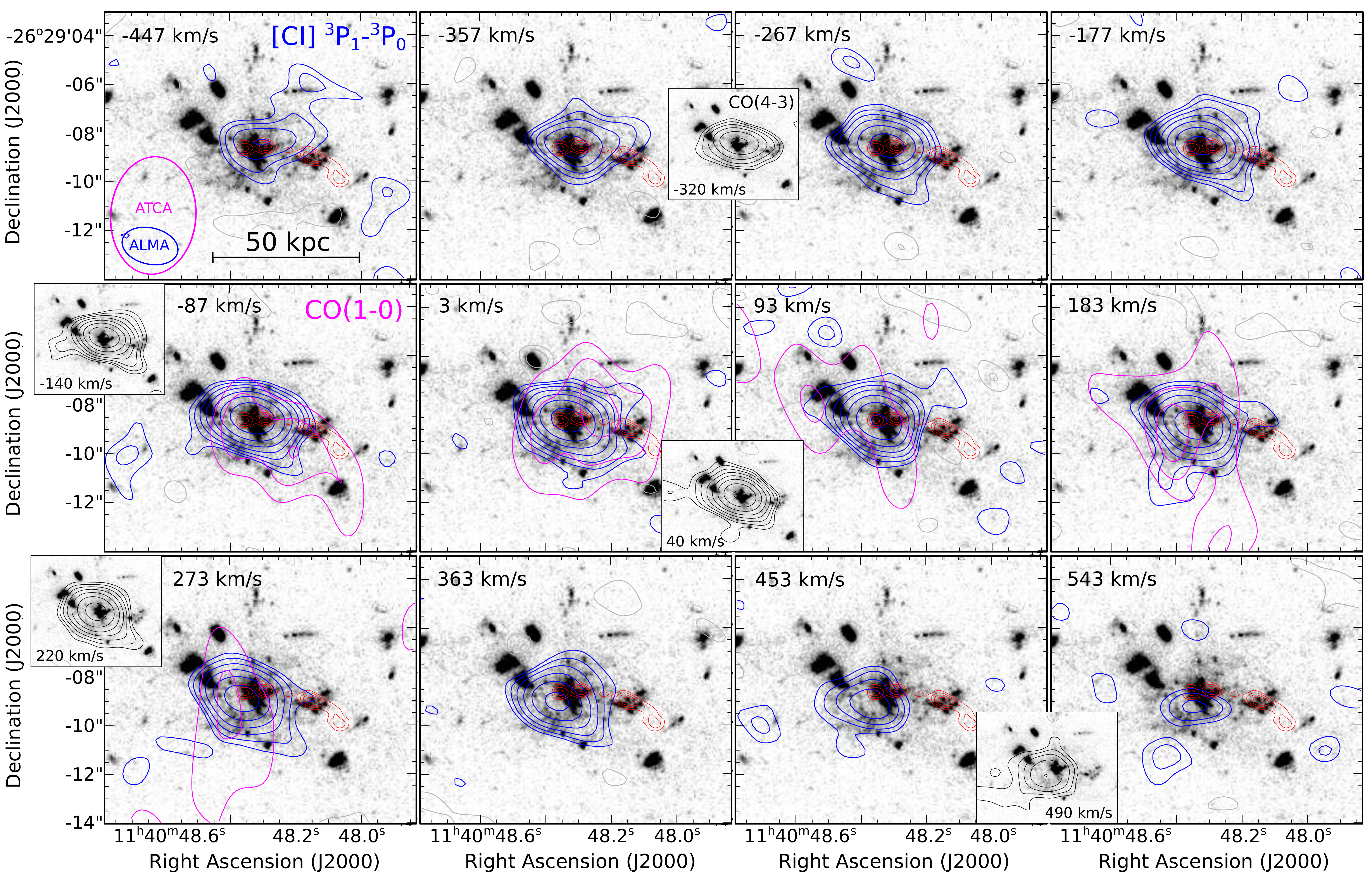}
\caption{Channels maps of the \cip\ emission (blue contours) over-plotted on an \textit{HST}/ACS F475W+F814W image of the Spiderweb Galaxy \citep{mil06}. The magenta contours indicate the previously detected CO(1-0) emission in channels where it is bright enough to be reliably detected (EM16). The most prominent features seen in \ci\ and CO(1-0) are also detected in CO(4-3) (black contours in the insets). All data-sets were binned to a velocity resolution of 90 \kms, and the central velocity of each channel is indicated. Contour levels of \ci\ and CO(4-3) start at 2$\sigma$ and increase by factor 1.5, with $\sigma$\,=\,0.07 mJy\,beam$^{-1}$ (negative contours are shown in grey). CO(1-0) contour levels are at 2, 3, 4, 5$\sigma$, with $\sigma$\,=\,0.086 mJy\,beam$^{-1}$. The red contours indicate the 36 GHz radio continuum (EM16). The synthesized beams of the ALMA and ATCA data are shown in the bottom-left corner of the top-left plot.}
\label{fig:IGM}
\end{figure*}

While the bulk of the \ci\ and CO(4-3) in the Spiderweb Galaxy is associated with the central radio galaxy MRC\,1138-262, we also detect emission across $\sim$50 kpc in the CGM (Fig.~\ref{fig:IGM}). As with our previous CO(1-0) results, the extended \ci\ is not co-spatial in either location or velocity with ten of the brightest satellite galaxies visible in Fig.\,\ref{fig:IGM} (\citealt{kui11}; EM16). The \ci\ and CO(1-0) appear to follow the same distribution and kinematics across the velocity range where both lines are reliably detected ($-$87 to 273 \kms\ in Fig.\,\ref{fig:IGM}). At the highest velocities, the \cip\ peaks $\sim$7~kpc SE of the core of the radio galaxy, at a location where previous high-resolution ALMA data found a concentration of \ci\ $^{3}$P$_{2}$-$^{3}$P$_{1}$ \citep{gul16}. The CO(4-3) and \ci\ show similar morphologies.

\begin{figure}
\centering
\includegraphics[width=0.37\textwidth]{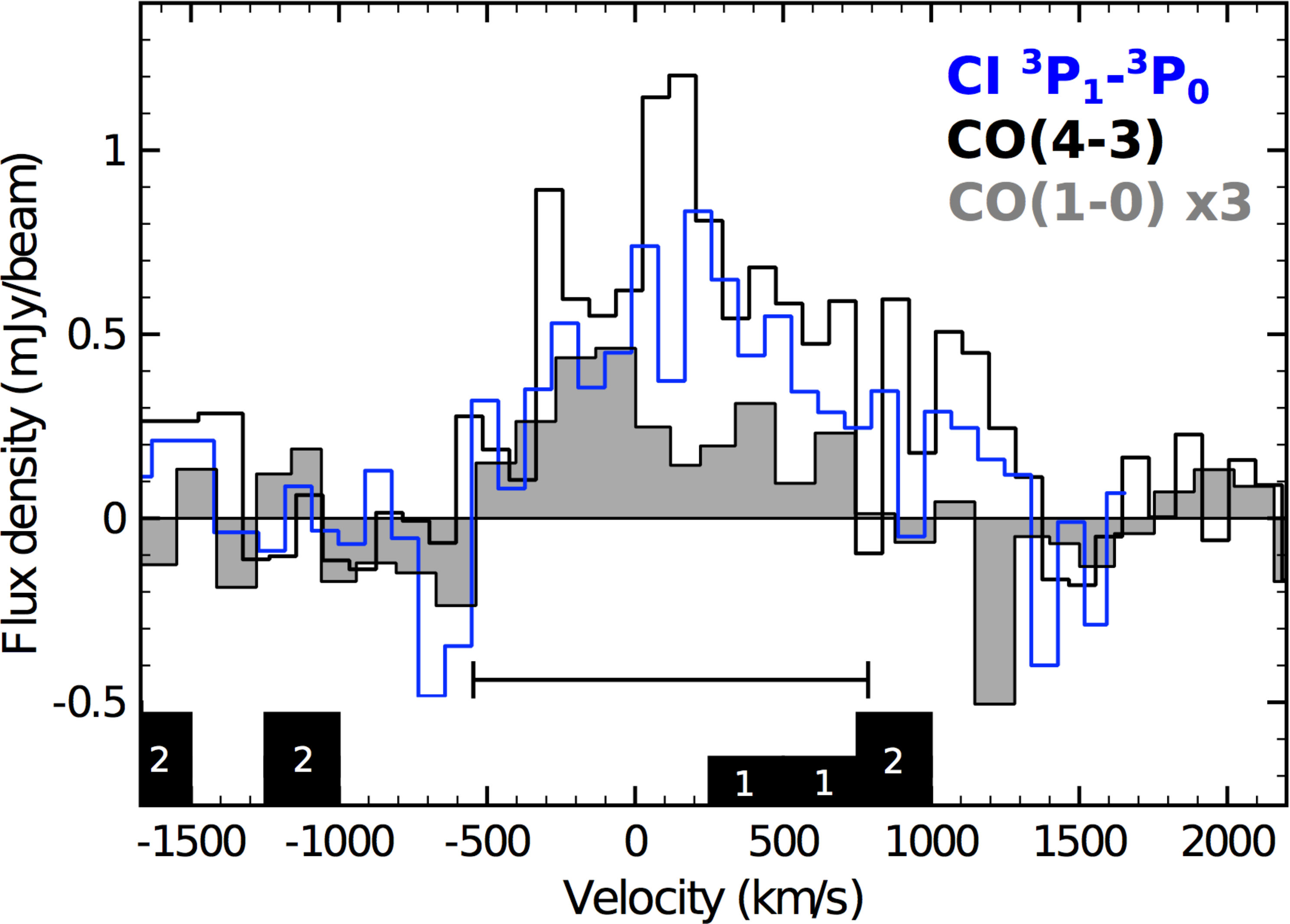}
\caption{Emission on 17-70\,kpc scales in the Spiderweb's CGM. The spectra were extracted by tapering the various data to $\sim$8$^{\prime\prime}$ and subtracting the central 2$^{\prime\prime}$ spectra of MRC\,1138-262 (Fig.\,\ref{fig:galaxies}). For the central CO(1-0) spectrum of MRC\,1138-262, we used the average between the untapered ATCA spectrum and a tapered high-resolution VLA spectrum from EM16, as explained in Table\,\ref{tab:results}.  The horizontal bar indicates the conservative velocity range over which we detect all three tracers in the CGM, which we used to determine intensities and ratios. The botton histogram shows velocities of satellite galaxies that lie within the molecular halo, based on [O\,II], [O\,III] and H$\alpha$ \citep{kui11}.}
\label{fig:spectraIGM}
\end{figure}

The bright \ci\ and CO(4-3) in the central $\sim$2$^{\prime\prime}$ ($\sim$17 kpc) beam make it non-trivial to determine flux densities across the CGM. We therefore taper the ALMA data to a beam of $\sim$8$^{\prime\prime}$ ($\sim$70 kpc), which covers the full CO(1-0) halo. We then take the spectrum within this tapered beam and subtract the line profile of the central radio galaxy (Fig.\,\ref{fig:galaxies}). The resulting spectra of the CGM are shown in Fig.\,\ref{fig:spectraIGM}. For both \ci\ and CO(4-3), $\sim$30$\%$ of the total flux is spread on 17-70 kpc scales (Table\,\ref{tab:results}). The ten bright satellite galaxies with known redshifts, and likely also any fainter satellites \citep{hat08}, do not substantially contribute to this emission. The reasons are that the galaxies have a much higher velocity dispersion than the gas \citep[Fig.\,\ref{fig:spectraIGM};][]{kui11}, and the 3$\sigma$ upper limit for even the brightest satellite galaxies is $I_{\rm [CI]} $<$ 0.028$ Jy\,beam$^{-1}$ $\times$ \kms\ (FWHM = 200\,\kms), or $<$5$\%$ of the \ci\ brightness of the CGM.

\section{Discussion}
\label{sec:discussion}

\begin{figure}
\centering
\includegraphics[width=0.38\textwidth]{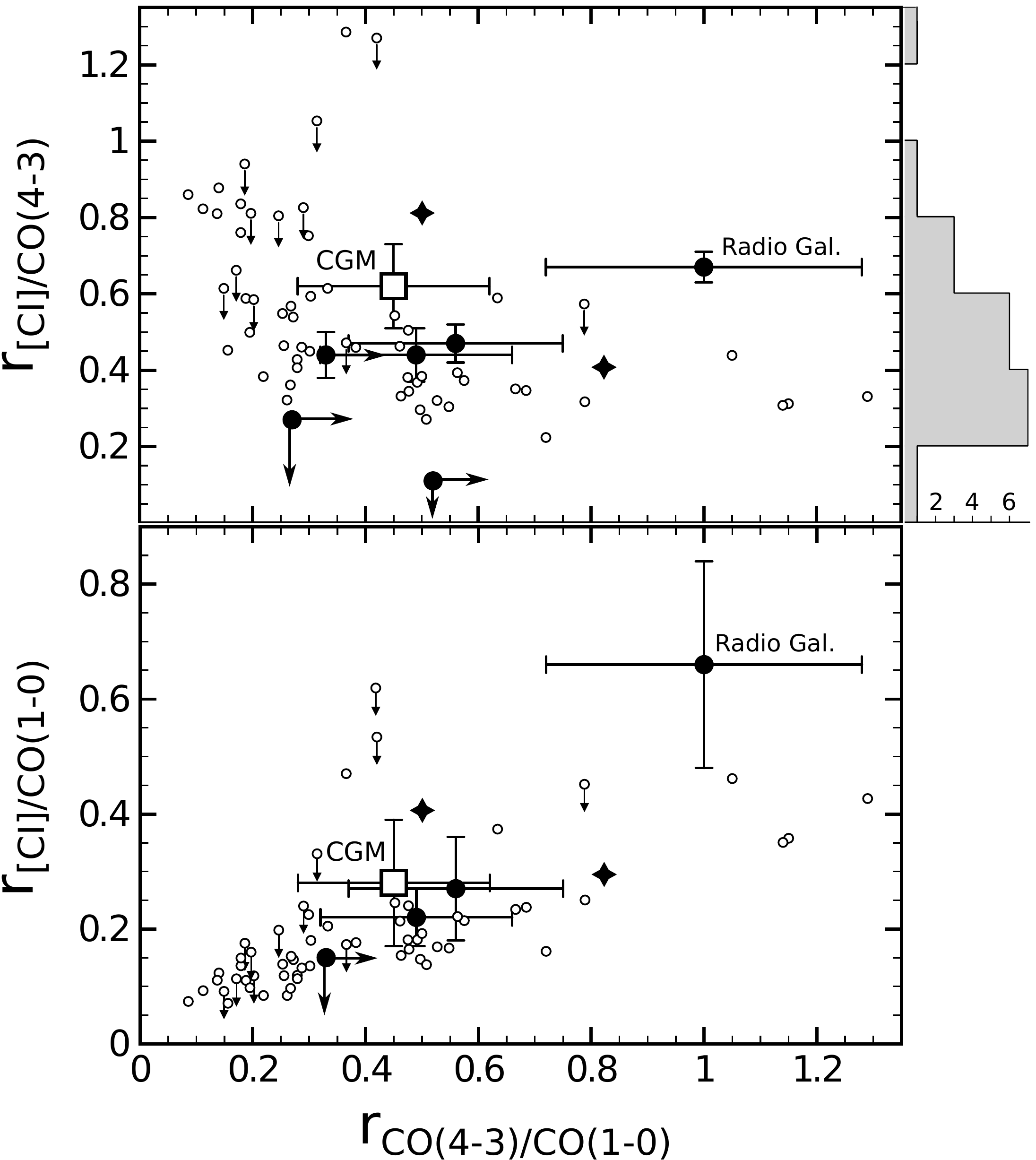}
\caption{Ratios of the \ci, CO(1-0) and CO(4-3) lines tracing a wide range of carbon abundances and excitation conditions. The open square represents the CGM of the Spiderweb Galaxy (17$-$70 kpc), the large solid dots the six proto-cluster galaxies from Fig.\,\ref{fig:galaxies}. The two stars represent two high-$z$ lensed SMGs \citep{dan11,les10,les11}, the small open circles low-$z$ star-forming galaxies \citep{kam16, isr15, ros15}. The histogram on the right ordinate of the top-panel is the r$_\mathrm{CI/CO(4-3)}$ distribution of high-$z$ SMGs and QSOs \citep{ala13,bot17}.}
\label{fig:ratios}
\end{figure}

Observations of \cip, CO(1-0), and CO(4-3) enable us to estimate the carbon abundance and excitation conditions of the molecular gas in the CGM and proto-cluster galaxies. Fig.\,\ref{fig:ratios} (top) shows that the values for $L'_{\rm [CI]}$/$L'_{\rm CO(4-3)}$ spread across a large range \citep[see also][]{wal11,ala13,bot17}. When instead comparing the ground-transitions of \cip\ and CO(1-0), Fig.\,\ref{fig:ratios} (bottom) shows two interesting results.  First, the CGM has excitation conditions, $L'_\mathrm{CO(4-3)}$/$L'_\mathrm{CO(1-0)}$, and relative \ci\ brightness, $L'_\mathrm{[CI]}$/$L'_\mathrm{CO(1-0)}$, similar to those of the proto-cluster galaxies, as well as low-$z$ star-forming galaxies. Second, both the gas excitation and relative \ci\ brightness are substantially higher in the radio galaxy  MRC\,1138-262. A possible explanation for the latter is that the CO(1-0) luminosity is reduced due to a high cosmic ray flux near the AGN \citep{bis17}. Alternatively, the \ci\ luminosity may depend on processes that also affect the gas excitation, and thus the luminosity of high-$J$ CO lines like CO(4-3) (Sect.\,\ref{sec:results}).

We estimate an H$_{2}$ mass in the CGM on 17$-$70 kpc scales of M$_\mathrm{H_2}$\,$\sim$\,1.0$\pm$0.4$\times$10$^{11}$($\alpha_\mathrm{CO}$/4)\,M$_{\odot}$ \citep[Table\,\ref{tab:results};][]{sol05}. The \ci\ mass in the CGM is M$_\mathrm{[CI]}$$\sim$8.9$\pm$1.4$\times$10$^{6}$\,M$_{\odot}$, assuming T$_\mathrm{ex}$$\sim$30\,K \citep{wei05}. This results in a \ci\ abundance of $X_\mathrm{[CI]}$/$X_\mathrm{H_2}$\,=\,M$_\mathrm{[CI]}$/(6M$_\mathrm{H_2}$) $\sim$ 1.5$\pm$0.6$\times$10$^{-5}$ (4/$\alpha_\mathrm{CO}$), close to that of the Milky Way ($\sim$2.2$\times$10$^{-5}$) and other high-$z$ star-forming galaxies \citep[][]{fre89,wei05,bot17}. The H$_{2}$ densities must be at least $\sim$100 cm$^{-3}$, which is the high end of densities of the cool neutral medium, where the \HI\ to H$_{2}$ transition occurs \citep{bia17}. More likely, values will be close to $\sim$\,500 cm$^{-3}$, the critical density of \ci\ \citep{pap04}.

\subsection{Mixing in the CGM}

Our findings of extended \ci, CO(1-0) and CO(4-3) imply that the cold molecular CGM is metal-rich and not diffuse. As we showed in EM16, the surface densities of the molecular CGM and the rate of in-situ star formation across the halo fall on the same Kennicutt-Schmidt relation as for star-forming galaxies \citep{ken98}. The fact that the gas excitation and \ci\ abundance of the CGM are similar to that of the ISM in star-forming galaxies strengthens this claim.

Despite the similarities between the CGM of the Spiderweb and the ISM in surrounding proto-cluster galaxies, it is unlikely that the cold CGM consists mainly of gas that is currently being tidally stripped from proto-cluster galaxies. If originally the gas was stripped, the low velocity-dispersion of the cold gas compared to that of the galaxies means that the gas must have had at least a dynamical time of t$_\mathrm{dyn}$\,$\gtrsim$\,10$^{8}$ yr to settle. Since the life-time of the OB stars across the CGM is only $\sim$10$^{7}$ yr, they must have formed long after the cold gas settled and cooled \citep[see][]{hat08}. 

Our results have important implications for our understanding of galaxy formation. Most importantly, the \ci\ and CO properties do not corroborate models of efficient and direct stream-fed accretion of relatively pristine gas \citep[e.g.,][]{dek09}. Instead, they agree with more complex models where the gas in the CGM is a melange from various sources -- metal-enriched outflows, mass transfer among galaxies, gas accretion, and mergers \citep{mor06,nar15, angl17, fau16}. If the gas becomes multiphase and turbulent as it flows, the interaction and mixing of gas from these various sources is likely efficient \citep[][]{corn17}. The gradual build-up of carbon, oxygen and dust that is starting to be modeled across these extended regions mimics many of the properties that we observe in the CGM of the Spiderweb. Thus, our results support the hypothesis that galaxies grow from recycled gas in the CGM and not directly out of accreted cold gas from the cosmic web. 

\section*{Acknowledgments}
We thank Padelis Papadopoulos for valuable feedback and expert advice on how to maximize carbon emissions and pollute the environments of galaxies. This paper makes use of the following ALMA data: ADS/JAO.ALMA$\#$2015.1.00851.S. ALMA is a partnership of ESO (representing its member states), NSF (USA) and NINS (Japan), together with NRC (Canada), MOST and ASIAA (Taiwan), and KASI (Republic of Korea), in cooperation with the Republic of Chile. The Joint ALMA Observatory is operated by ESO, AUI/NRAO and NAOJ. The Australia Telescope is funded by the Commonwealth of Australia for operation as a National Facility managed by CSIRO. The National Radio Astronomy Observatory is a facility of the National Science Foundation operated under cooperative agreement by Associated Universities, Inc. This research received funding from the Spanish Ministerio de Econom\'{i}a y Competitividad grants Ram\'{o}n y Cajal RYC2014-15686 (HD) and AYA2015-64346-C2-2-P (MV) and the Australian Research Council Centre of Excellence for All-sky Astrophysics in 3D project CE170100013 (JA).\\
\vspace{-10mm}\\

\end{document}